# *h/e* Oscillations in Interlayer Transport of Delafossites

Carsten Putzke[1*], Maja D. Bachmann[2,3], Philippa McGuinness[2,3], Elina Zhakina[2], Veronika Sunko[2,3], Marcin Konczykowski[5], Takashi Oka[2,4], Roderich Moessner[4], Ady Stern[5], Markus König[2], Seunghyun Khim[2], Andrew P. Mackenzie[2,3*], Philip J.W. Moll[1*]

[1]ILaboratory of Quantum Materials (QMAT), Institute of Materials, École Polytechnique Fédéral de Lausanne (EPFL), 1015 Lausanne, Switzerland
[2]Max Planck Institute for Chemical Physics of Solids, 01187 Dresden, Germany
[3]School of Physics and Astronomy, University of St. Andrews, St. Andrews KY16 9SS, UK
[4]Max Planck Institute for the Physics of Complex Systems, 01187 Dresden, Germany
[5]Weizmann Institute of Science, Departement of Condensed Matter Physica, Rehovot, Israel
[6] Laboratoire des Solides Irradiés, CEA/DRF/IRAMIS, _ Ecole Polytechnique,CNRS, Institut Polytechnique de Paris, F-91128 Palaiseau, France

*To whom correspondence should be addressed: carsten.putzke@epfl.ch, andy.mackenzie@cpfs.mpg.de, philip.moll@epfl.ch

**Transport of electrons in a bulk metal is usually well captured by their particle-like aspects, while their wave-like nature is commonly harder to observe. Microstructures can be carefully designed to reveal the quantum phase, for example mesoscopic metal rings resembling interferometers. Here we report a new type of phase coherent oscillation of the out-of-plane magnetoresistance in the layered delafossites $PdCoO_2$ and $PtCoO_2$. The oscillation period is equivalent to that determined by the magnetic flux quantum, *h/e*, threading an area defined by the atomic interlayer separation and the sample width. The phase of the electron wave function in these crystals appears remarkably robust over macroscopic length scales exceeding 10μm and persisting up to elevated temperatures of T>50K. We show that, while the experimental signal cannot be explained in a standard Aharonov-Bohm analysis, it arises due to periodic field-modulation of the out-of-plane hopping. These results demonstrate extraordinary single-particle quantum coherence lengths in the delafossites, and identify a new form of quantum interference in solids.**

Electrons in vacuum carry the characteristics of particles as well as waves, which is demonstrated in interference experiments directly probing the phase information[1]. In metals the transport properties are usually well captured by the particle nature of the electron only, summarized in the semi-classical Boltzmann equation. The wave-like character is masked by the high density of electrons and their interaction with the ionic lattice, which leads to a loss of the phase information in bulk phenomena. With experimental effort, samples can be fabricated on the mesoscopic length scale over which the phase of the electron is preserved, thus becoming observable in electronic transport. A well-known example is the Aharonov-Bohm effect (ABE) in nanoscopic rings of gold[2,3], which presents a solid-state analog of the interference experiment by Davisson and Germer[1]. Common to these experiments is the creation of an artificial loop enclosing magnetic flux which acts as a beam splitter.

The main experimental observation we report here is a novel and surprisingly robust manifestation of phase coherence intrinsic to the out-of-plane transport in single bars of the ultra-pure delafossites $PdCoO_2$ and $PtCoO_2$. These materials are composed of highly conducting Pd/Pt layers separated by $CoO_2$ layers, reflected in a large transport anisotropy $\rho_c/\rho_a$ exceeding 1000. The layered triangular crystal lattice leads to an almost hexagonal Fermi surface (FS)[4] with little warping, which has been well characterized by de Haas-van Alphen oscillations[5] and angle-dependent magnetoresistance oscillations[6,7]. These materials are the most conductive oxides known, with an in-plane transport mean free path (mfp) of more than 20μm at low temperatures[5,8,9].

The strong anisotropy is also reflected in the growth of thin plate-like crystals, a common property of layered materials. While mesoscopic quantum phenomena are successfully probed in the plane of quasi 2D ultra-pure metals, achieving such electrical transport perpendicular to the layers is challenging. We have overcome this difficulty by employing focused ion beam (FIB) micro-structuring techniques[10]. Starting from as-grown crystals, we have milled pillars along the c-axis, thereby restricting the in-plane dimensions to few micrometers. A typical structure designed for four-point resistivity measurements is shown in figure 1A. Since the depth $d$ and width $w$ of the pillar are both well below the mfp, the system enters the ballistic transport regime in the plane.

Applying an in-plane magnetic field at low temperatures, we find an oscillatory magnetoresistance (figure 1B). These oscillations are clearly visible in the raw data ($\Delta\rho_{osc}/\rho \sim 5\%$, Figure 1). To perform further analysis, we focus on the second derivative of the magnetoresistance (figure 1C). The oscillations are periodic in magnetic field and their periodicity scales inversely with the width of the pillars over an order of magnitude, from 1.2µm to 12µm (figure 1D).

The observed periodicity matches remarkably well with that expected for a magnetic flux quantum $\Phi_0=h/e$, with the Planck constant $h$ and electron charge $e$, threading through an area $S$ enclosed by two adjacent Pd/Pt layers and the sample side walls (dashed line in figure 1D). This gives an area $S = w^*c/3$, where $w$ is the width of the sample and $c$ denotes the crystallographic unit cell (PdCoO$_2$: 1.774nm; PtCoO$_2$: 1.781nm)[11]. Due to the ABC stacking the unit cell contains three Pd/Pt layers, hence the relevant height is $c/3$. Such oscillations of the magnetoresistance, periodic in $\Phi_0$, demonstrate quantum transport of coherent electron waves spanning the width of the entire sample. Oscillations are readily observed in samples as wide as 12 µm, thus indicating a macroscopic phase preservation in the metal. PdCoO$_2$ has been shown to have an extremely long ballistic mfp[12], but observation of phase coherence in a high carrier density metal over such a long distance is still surprising. It is particularly noteworthy that no special care to decouple the sample from the environment had to be taken, such as ultra-low temperatures or special substrate decoupling (all samples are simply attached to a sapphire chip by epoxy glue, see methods). Still, all samples from different crystals show a highly consistent picture of strong long-ranged quantum coherence.

So far, we have only considered magnetic fields applied perpendicular to the sample surface. If indeed the oscillation frequency is set by the flux through the area $S$, it would be natural to expect a sinusoidal dependence on the magnetic field angle when rotating within the Pd/Pt layer. The experimental frequency spectrum upon rotation is more complex, with multiple frequencies appearing (see figure 2). A natural geometric interpretation of the angle dependence is found because of the FS topography of PdCoO$_2$. The almost perfect hexagonal FS, in contrast to a circular one, exhibits three preferential directions of electron motion perpendicular to the flat faces of the FS. In real space this describes three interweaving subsystems of directional electron flow in the plane, each spanning its own area $S_i$ (i=1..3, sketched in figure 2A). The flux enclosed in each subsystem contributes oscillations of frequency $|\vec{B} \cdot \vec{S}_i|/\Phi_0$ to the total conduction, leading to three, 60° offset, branches in the frequency spectrum. The difference in symmetry between the hexagonal FS and the rectangular sample shape divides the branches into two different types. The samples are cut such that one preferential direction of motion is aligned with a sample side wall. Therefore, one subsystem area is set by the full sample depth, while two symmetric branches are related geometrically to the sample width. The aspect ratio of the cross section is reflected in the relative ratio of the maximum frequency values in the two types of branches. From the in-plane angle dependence it follows that for the magnetic field aligned with both the sample sides ($\theta = 0°$ and 90°) an area $S_i$ scales with $w$ or $d$ respectively. In figure 1D both angle configurations were combined by denoting the dimension perpendicular to magnetic field as sample width $w$.

The period of all field-induced oscillations in quantum objects is given by an integer multiple of the flux quantum threading through them, $B_n \cdot S = n\Phi_0$. Usually, the relevant length scale in metallic systems is set by the magnetic field itself, in form of the cyclotron radius $r_c$, leading to oscillations periodic in $1/B$ ($B_n \cdot r_c^2 \propto n\Phi_0$). The most famous of such $1/B$ periodic magnetoresistance oscillations are Shubnikov-de Haas oscillations[13]. Given the 2D nature of the FS, SdH oscillations will not appear for in-plane fields as all orbits are open. However, as the magnetic field is rotated out of the Pd-layers, the out of plane field induces orbital motion and will lead to usual SdH oscillations. This leads to a particularly rich interplay between the different quantum transport regimes as a function of out-of-plane angle $\gamma$ (figure 3). For an in-plane magnetic field ($\gamma=0°$) $B$-periodic oscillations are observed as previously discussed. On tilting the field out of plane, those oscillations are limited to lower magnetic fields and vanish at a field scale $B^*$ defined by the mesoscopic size of our samples. $B^*$ corresponds to the angle-dependent field scale required to fit a bulk-like cyclotron radius into the pillar, given by the condition of $2r_c=w$. $B^*$ appears as a clear anomaly in the magnetoresistance, delineating a strong negative magnetoresistance above $B^*$. Thereby, tilted magnetic fields induce a transition between mesoscopic quantum transport in the low field regime and bulk-like transport described by Landau levels at sufficiently high fields. This picture is naturally apparent: Once the in-plane the Lorentz force is sufficient to bend a wave front back on itself, it will self-interfere leading to Landau quantization. Naturally, this detaches the wave function from the boundary, hence the Landau levels are entirely bulk-like and independent of sample width. This scenario is further supported by the negative magnetoresistance above $B^*$, as the dominant boundary scattering in the clean devices is suppressed as bulk-like Landau tubes form in the core of the pillar. Indeed, above $B^*$ conventional SdH oscillations are observed in our samples, which coexist with $B$-periodic oscillations in the intermediate angle range. The SdH frequencies and effective masses of $m^*\approx 1.5 m_e$ (PdCoO$_2$[5]) and $1.2 m_e$ (PtCoO$_2$[7]) are consistent with work performed on macroscopic crystals, clearly excluding the possibility that the microfabrication has strongly altered the material. The large size of the hexagonal Fermi surface leads to high-frequency oscillations around F∼30kT, which are well observed in the microstructure (see supplement). However, resolving such high frequencies requires very slow field sweeps (<10mT/min), which are impracticable to perform over large field ranges and multiple angles. Therefore, only the slow difference frequency corresponding to the beating of neck and belly frequencies is apparent in figure 3, yet the main frequencies were always observed consistently when sweeping more slowly.

Further insights into the quantum transport arise from a comparison of the transport and quantum mean free paths. The quantum coherence length extracted from SdH oscillations is found to be only 400nm (see supplement), more than an order of magnitude smaller than that observed in the $B$-periodic oscillations. Furthermore, the quantum coherence length remains almost unchanged in irradiated samples, while the in-plane transport mean free path is reduced by more than a factor 10. It is important to consider that the quantum mean free path obtained from a Dingle analysis represents an average over the entire FS orbit, while the B-periodic oscillations are exclusively due to the flat sections of the hexagonal FS. A resolution to this conundrum would be a large quantum scattering rate at the corners of the hexagon; evidence of this has previously been reported in an analysis of the Hall effect[14].

B-periodic oscillations, such as those reported here, arise when a field-independent area S enters the quantization condition, such that $B_n \cdot S = n\Phi_0$. This is very rare in metals due to a mismatch of length scales. On the small side, any area associated with interatomic distances within the unit cell is too small, because the observation would require extreme fields to achieve a flux quantum ($\Phi_0/\text{Å}^2 \sim 10^5$T). On the large side, the finite quantum coherence length usually prevents the electron wave from spanning the entire sample dimension. The few known $B$-linear oscillatory phenomena in solids are either non-quantized or rely on a superconducting order parameter to establish macroscopic phase coherence or exploit artificially introduced nanometric length scales. Such examples include semi-classical resonances (Sondheimer[15]), interlayer superconducting flux quantization (intrinsic Josephson junctions[16]), cyclotron motion in thin films (Azbel-Kaner[17]) and interference between parallel quantum wells[18]. None of these can explain the here-reported magnetoresistance oscillations as the $h/e$-periodicity clearly indicates long ranged single particle phase coherence as their origin.

Given the flux quantisation condition that we have identified, it is at first sight appealing to invoke a scenario akin to the Aharonov-Bohm effect (ABE) to account for our data. In this picture, the quasiparticle would encircle the area S with the Pd/Pt-layers resembling the arms of an interferometer. Since the nanoscopic dimensions of the interlayer distance would be combined the macroscopic sample width, the necessary fields would be scaled to the range accessible in superconducting magnets. While this would naturally lead to the observed periodicity, this scenario has severe shortcomings. For one, the Pd-layers are too strongly coupled via the sizable out-of-plane hopping element $\tau_\perp$ as determined by quantum oscillations ($\tau_1=1$eV; $\tau_\perp=10$meV)[8] leading to an infinite number of paths and destructive interference. Furthermore, the ABE is most commonly accompanied by a related self-interference due to weak localization, the Al'tshuler-Aronov-Spivak (AAS) effect[19] giving oscillations periodic in $h/2e$ in metallic rings[2,3,20]. This, or any other higher order quantum process involving multiple layers in a stack ($2S, 3S,...$), are not observed experimentally here. If present, their small amplitude is hidden within the noise level of

1% of the h/e oscillations (see supplement). A third key feature of our observation is its robustness to temperature (Fig. 4). The observation of the ABE in metallic rings was limited to below 1K due to decoherence from coupling to modes in the substrate, something that is also a feature of reported ABE experiments on graphene[21]. This is in stark contrast to the T>50K temperature scale observed here in highly metallic PdCoO$_2$ and PtCoO$_2$. Similar high-temperature quantum coherence has been seen in bismuth nano-wires[22], topological nano-ribbons[23,24], quantum dots[25], carbon nano-tubes[26] and 1/B superlattice oscillations in graphene[27] but the key to its observation was the small length scale involved in those nano-scale systems. This is not surprising in light of the calculation presented in the Supplementary Online Material, in which we show that even if we impose a far larger $\tau_\perp$ at the sample edge than in the bulk, an ABE-style calculation predicts an experimental signal dying out as $1/w$. For our situation in which w is approximately $10^4$ lattice spacings, it would be unresolvable.

The limitation to length scales smaller than the mfp should also be reflected in the onset temperature of the oscillation. As the mfp decreases with increasing temperature one would expect the h/e oscillations to persist up to higher temperatures for samples with smaller width w. In contrast the onset temperature appears independent of sample width, which suggests that the upper temperature limit is not set by the mfp in the sample. In order to further probe this observation, we performed an additional set of measurements on a device whose mfp had been reduced by a factor of 20 to 1μm by the introduction of point defects created by 2.5 MeV electron irradiation[28]. Oscillations were unresolvable when the device width was 8μm and reappeared when it was narrowed to 1μm (see supplementary figure 5), in line with the expectation that the mfp must be of order of the sample width or larger for the signal to be seen. Strikingly, the temperature dependence of the oscillations in the disordered device is the same as that in the cleaner devices within experimental error (see Fig. 4B). This strongly suggests that if the zero temperature in-plane mfp exceeds the sample width, the oscillatory signal persists until $k_BT$ becomes a substantial fraction of $\tau_\perp$ despite the mfp becoming smaller than the device width at elevated temperatures.

This leads to a different scenario, providing what we believe to be a very promising framework for understanding our observations. In this, the purity of the experimental system plays an intriguing dual role. Firstly, it provides a wavefunction with full phase coherence across the length of the ab plane. Secondly, the microscopically regular structure of the essentially perfect delafossite crystals gives rise to a periodic array of tunnelling paths between the layers in the c direction.

In our picture, a phase coherent wave in the ab plane is transmitted to an adjacent plane with tunnelling matrix elements $\tau_\perp << \tau_1$ in each unit cell of the lattice. For a field applied along b, the phase of the tunnelling matrix elements at site j=1...L is modulated by a factor $e^{i 2 \pi \frac{\varphi}{L} j}$. The field strength is written such that there are $\varphi = |\vec{B} \cdot \vec{S}_i|/\Phi_0$ flux quanta per layer across the system of width L (=w/lattice constant) in the a-direction. Summing the resulting series yields

$$A(\varphi) = \sum_{j=1}^{L} e^{i 2 \pi \frac{\varphi}{L} j} = \frac{1 - e^{i 2 \pi \varphi}}{e^{-i 2 \pi \frac{\varphi}{L}} - 1}.$$

The reader will recognise this as equivalent to the far-field diffraction pattern from a diffraction grating of finite width L illuminated by a coherent light field. As is well known, the finite width of such a grating is modelled by multiplying the transmission function of an infinite grating with a boxcar function (also known as a top-hat function), rect(L). Invoking the convolution theorem of the Fourier transformation, its far-field diffraction pattern becomes convolved with the Fourier transform of rect(L) namely sinc(L). This is, we believe, the key physics behind our observation, because this convolution gives rapid oscillations at a frequency that depends on L but with a relative amplitude from one oscillation to the next that is approximately L-independent. In our experiment, the applied field is small such that $\frac{\varphi}{L} \ll 1$ and we observe only the first few oscillations of the sinc function.

Reflecting this intuitive picture, we have performed a transport calculation using the finite system Kubo formula[29–31] in the anisotropic Hofstadter model, which is summarized in Figure 5 and presented in supplementary information. As can be seen in Fig. 5, it reproduces the qualitative experimental signal very well, with the expected sinc function clearly evident in Fig. 5C. It also provides a further key insight. The bandwidth for interplane transport (Fig. 5B) is seen to be modulated by the field, even vanishing for special field at which $\varphi$ is an integer, even though the bare hopping $\tau_\perp$ remains unchanged.

The extremely anisotropic nature of the area pierced by the flux quantum -- truly microscopic in one direction, almost macroscopic in the other -- is a most striking reflection of the regime of electronic behaviour accessible in our compounds. This represents, to the best of our knowledge, a new, experimentally accessible regime of the famous Hofstadter Hamiltonian, in which the hopping terms in the plaquette are highly anisotropic.

A natural question is why these oscillations, seen so clearly in our raw data, have not been observed before. The answer is that, in addition to the rarity of crystalline perfection at the level found in the delafossites, only recent technological advances have enabled experimental investigations of this regime[10]. The technical key to this observation is to reduce $w$ to a few µm as well as to shape the sample cross-section into an ideal rectangle by FIB. This acts as a magnifying glass that allows us to venture deeper into the mechanism behind coherent inter-layer transport.

We believe that the observations and analysis reported in this paper will stimulate further experimental, theoretical and technological research. The framework that we have presented for understanding the data invites refinement, and similar physical picture might be developed from slightly different starting viewpoints. It may also be possible to extend the experiments to other high purity layered compounds such as the ruthenates. Furthermore, the evolution of the signal could be studied by fabricating bi- and few-layer thin films. Such thin films may also be used to explore technological possibilities. As quantum coherence emerges as its own subject in technology, it will be interesting to explore if novel applications can exploit the rare macroscopic single-particle phase coherence in the delafossites. The observation of this striking phase coherent process in a well-studied, single-band metal may prove to be a prime example of unknown phenomena as well as novel functionalities hiding in seemingly well-understood condensed matter systems.

**Acknowledgements** We thank Brad Ramshaw, Roni Ilan, Mark Fischer, Alberto Morpurgo and Laszlo Forro for helpful discussion. This project was supported by the Max-Planck Society and has received funding from the European Research Council (ERC) under the European Union's Horizon 2020 research and innovation programme (grant agreement No 715730) and the Deutsche Forschungsgemeinschaft (DFG, German Research Foundation) – MO 3077/1-1, as well as the Würzburg-Dresden Cluster of Excellence on Complexity and Topology in Quantum Matter (EXC 2147). M.D.B., P.M. and V.S. acknowledge studentship funding from EPSRC under grant no. EP/L015110/1. AS was supported by the Israel Science Foundation, by the European Research Council (Project LEGOTOP) and by the DFG through project CRC-183. MK acknowledges support from SIRIUS irradiation facility with project EMIR 2019 18-7099.

**Author Contributions** C.P., M.D.B., P.M., E.Z and M.K. fabricated the microstructures. V.S., E.Z., P.M. and M.K. performed the electron irradiation. Single crystals were grown by S.K. Transport measurements and data analysis were performed by C.P. and P.J.W.M. P.J.W.M., C.P., T.O., R.M. A.S. and A.P.M. worked on the interpretation of the data. All authors were involved in the design of the experiment and writing of the manuscript.

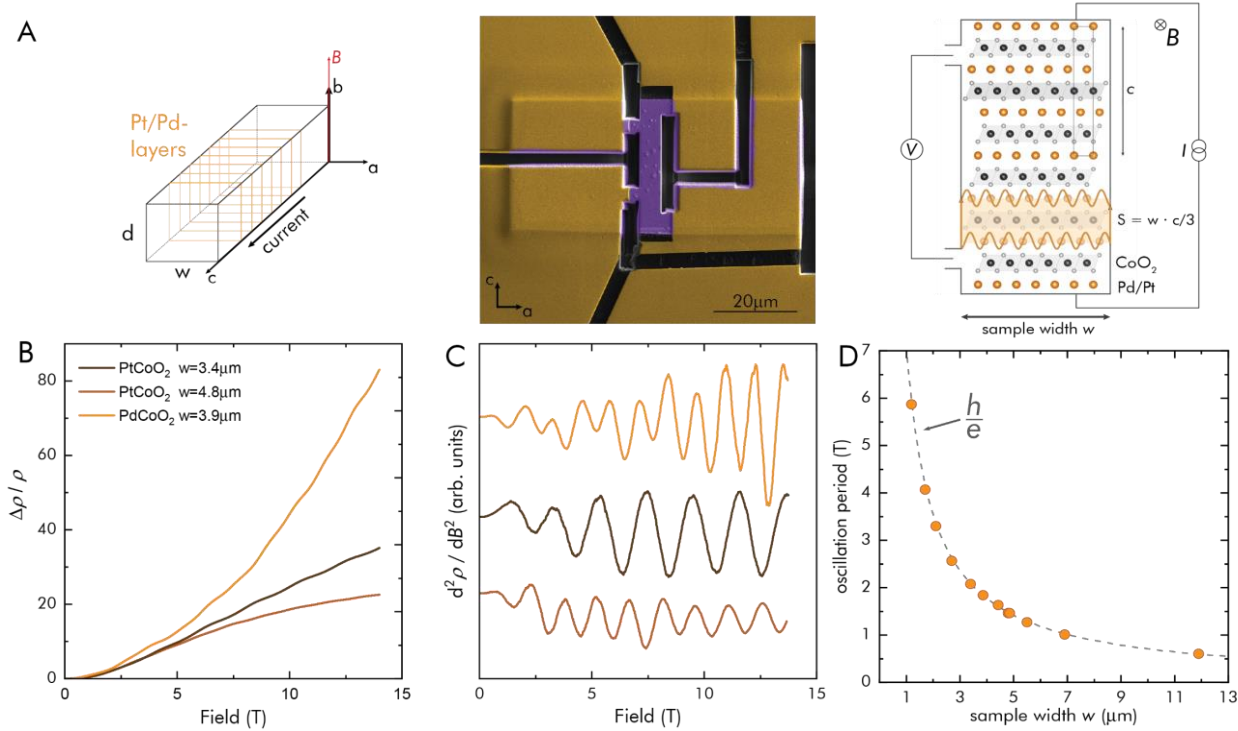

**Figure 1: Magnetoresistance oscillations periodic in magnetic field**
A) Experimental setup: (left) Current is passed along a bar shaped sample, perpendicular to the layered structure. The bars have a width $w$ and a thickness $d$. Magnetic field is applied and rotated within the Pd/Pt-layer. (middle) SEM image of $PdCoO_2$ microstructure to measure c-axis resistivity. (right) Crystal structure of $PdCoO_2$. Alternating layers of Pd/Pt and $CoO_2$ lead to a high anisotropy of the resistivity. The area $S$ relevant for the $h/e$-oscillations is spanned by two adjacent Pt/Pd layers.
B) Magnetoresistance of $PtCoO_2$ and $PdCoO_2$ at $T=2K$ of various sample widths for fields along the a-axis. The apparent difference in the high field background is due to a sharp feature in the angle dependent magnetoresistance when fields are close to parallel with the Pd/Pt-layers[7].

C) The second derivative of the resistivity highlights the oscillatory part of the magnetoresistance in panel *B*. Multi-frequency components are well explained by small sample misalignment (see figure 2).
D) The oscillation period is shown for different sample widths. The sample width dependence shows a remarkable agreement with the oscillation period expected for a single particle magnetic flux quantum, $h/e$, per area $S = w*c/3$ (as indicated in the panel above).

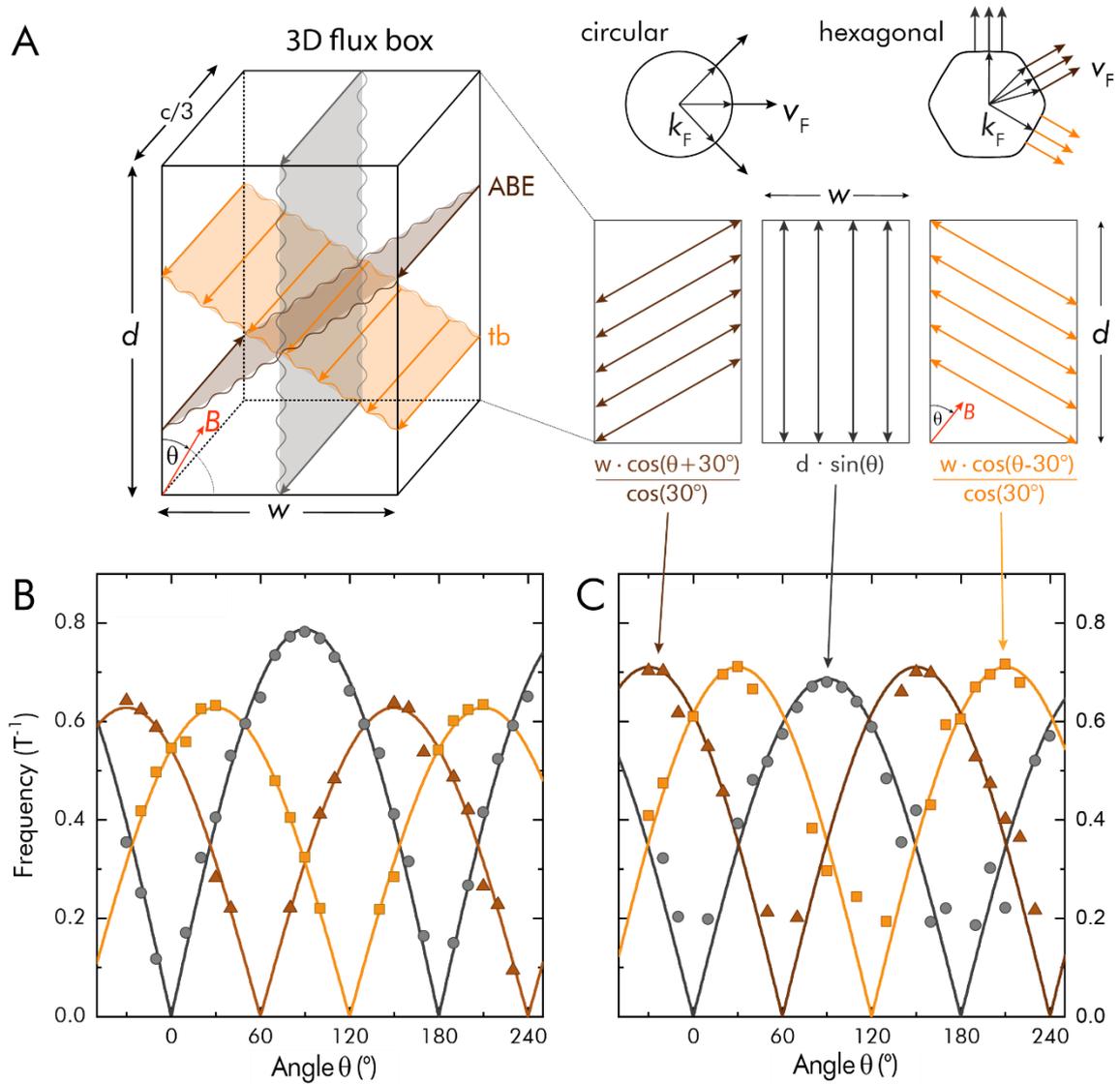

**Figure 2: Angle dependence of quantum coherent oscillations.**
A) Pd/PtCoO$_2$ possess almost hexagonal Fermi surfaces[4,32]. This leads to three preferred directions of motion in contrast to the case of a circular Fermi surface. The magnetic field is rotated in the planes. The three ballistic paths and their angle dependent projections on the magnetic field are shown. In rectangular samples two symmetric branches are set by the sample width and one by the sample thickness $d$ (see discussion in main text). The left panel illustrates the relevant 3D flux box limited by the sample width, thickness and two adjacent Pd/Pt-layers. This box defines the flux surfaces. The oscillations are periodic in integer flux quanta threading through them. B+C) Angle dependence of the quantum coherent oscillations of PdCoO$_2$ (B) and PtCoO$_2$ (C). Solid symbols represent the measured data points, solid lines show the expectation from the model sketched in panel A. B) shows data from a sample with a $d/w$-ratio of 1.4, while the ratio for the sample in panel C is close to 0.9.

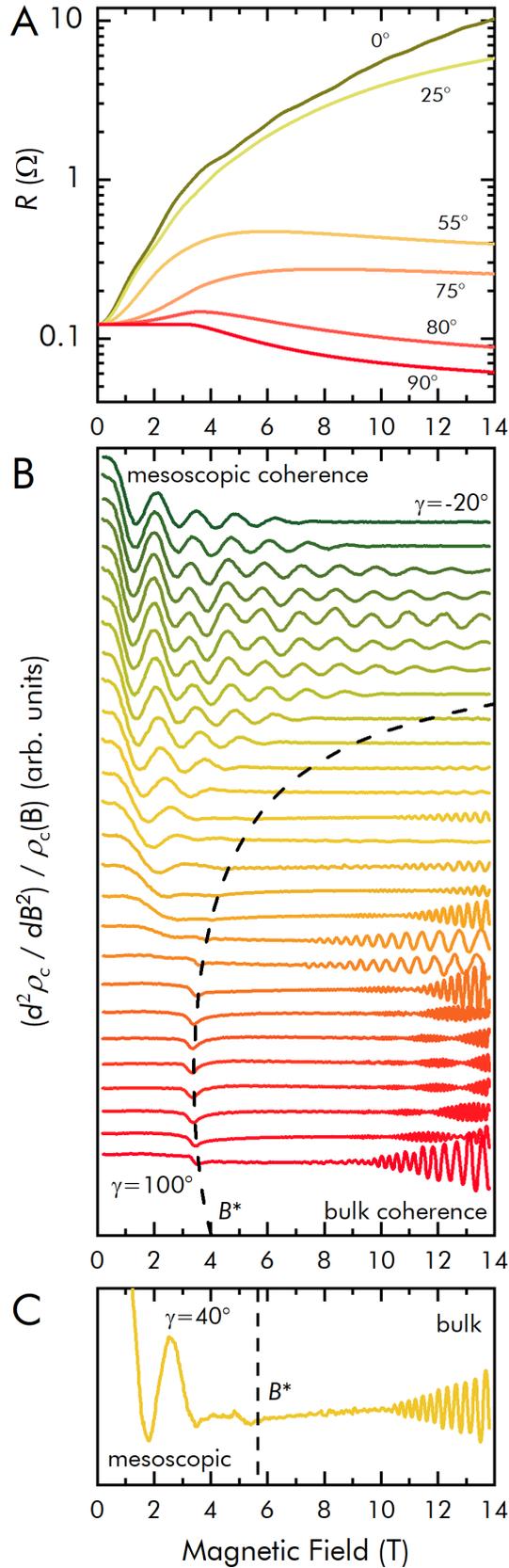

**Figure 3: Angle dependence tilting the field out of the plane.**
A) Magnetoresistance of PdCoO$_2$ (w=3.9µm) at different angles $\gamma$ between 0° (magnetic field in the plane) and 90° (magnetic field perpendicular to the layers). A negative magnetoresistance is observed for fields above 3T. At this field the cyclotron diameter 2r$_c$ becomes smaller than the sample width w (figure S3).
B) Second derivative of the magnetoresistance with respect to the magnetic field. For angles $\gamma$=0° B-periodic oscillations in agreement with figure 1 are shown. As the field is tilted out of the Pd-layers in 5° steps the oscillation period is modified as 1/cos $\gamma$. At higher tilt angles the B-periodic oscillation vanish and SdH oscillations are observed. Detailed analysis of these is shown in supplementary figure 4. The dashed line represents the field B* at which the cyclotron diameter coincides with the sample width w. B-periodic oscillations are seen over a wide-angle range below B*, while SdH oscillations only appear above B*. The data have been offset proportional to the magnetic field angle.
C) Subset of the data in panel B at a magnetic field angle of $\gamma$=40°. At low field the B-periodic oscillations are seen, while at high field 1/B-periodic oscillations are observed.

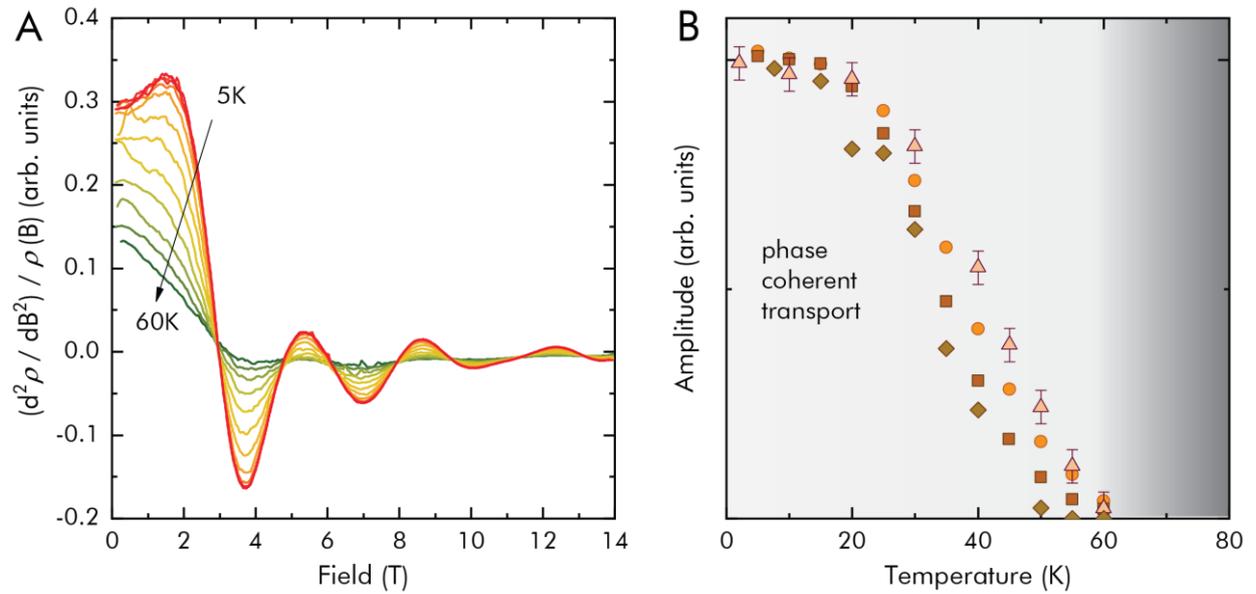

Figure 2: Temperature dependence of the phase coherent oscillation amplitude.
A) Second derivative of the magnetoresistance of $PtCoO_2$ (w=2.0μm). Oscillations persist up to 60K.
B) Oscillation amplitudes extracted from Fast Fourier Transform analysis in the field range from 3T to 12T are shown for $PtCoO_2$ (w=4.8μm; squares and w=2.0μm; circles), and $PdCoO_2$ (mfp 20μm, w=3.9μm; diamonds and mfp 1μm, w=1μm; triangles). In spite of the large changes of width and mfp, the temperature dependence of the signals from different samples is very similar, and the oscillations can be observed to remarkably high temperatures of > 50K.

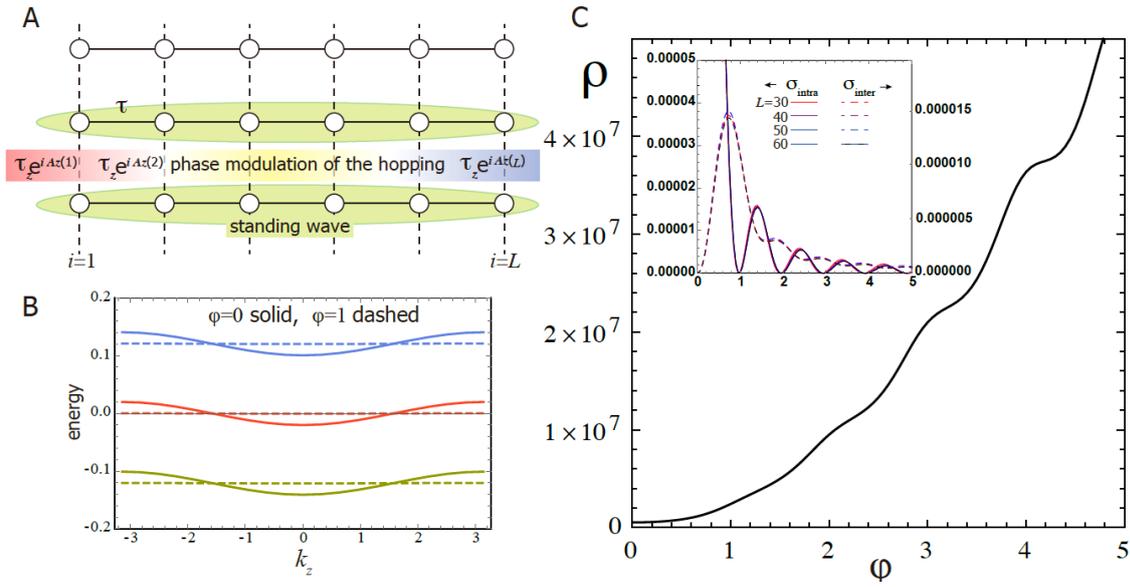

**Figure 5: Transport analysis for the anisotropic Hofstadter model.**
A) The anisotropic Hofstadter model with tunnelling parameters $\tau_\parallel = 100\ \tau_\perp = 1$ as a toy model to study the origin of the $h/e$ magneto-oscillation shown in Fig. 1. Interlayer transport occurs as an electron in one layer, in an extended standing wave state, tunnels to an adjacent layer. In the presence of a magnetic field, the tunneling matrix element is modulated by a phase factor $e^{i 2\pi \frac{\varphi}{L} j}$. This model naturally yields a vanishing interplane bandwidth, *given by $2\tau_\perp Re\ A(\varphi)$ in the large L and anisotropic limit,* for integer values of $\varphi$.
B) The energy spectrum of the anisotropic Hofstadter model around zero energy for L=51 showing vanishing bandwidth at $\varphi = n$ (integer).
C) Resulting calculations of the interlayer resistivity. Full details of the calculation are presented in supplement.. In the absence of incoherent (e.g. phonon assisted) interlayer processes, the resistivity would diverge for integer $\varphi$ since only intraband term $\sigma_{intra}$ in the Kubo formula contributes (inset). However, for large $L$ the level separation of Fig. 5B becomes smaller than the level broadening from such incoherent processes, motivating the inclusion of interband contributions $\sigma_{inter}$ in the calculation whose results are shown in Fig. 5C.

# Supplementary Information

**Crystal growth.** Single crystals of $PdCoO_2$ were grown in the evacuated quartz ampule with a mixture of $PdCl_2$ and CoO by the methathetical reaction : $PdCl_2 + 2CoO \rightarrow 2PdCoO_2 + CoCl_2$[1]. For $PtCoO_2$, $PtCl_2$ and CoO were used correspondingly. The ampule was heated at 1000°C for 12 hours and stayed at 700-800°C for 5-20 days. In order to remove excess $CoCl_2$, the resultant product was washed with distilled water and ethanol.

**Micro-structuring.** Transport lamellae perpendicular to the Pd/Pt layers were cut from as grown high quality single crystals using a FEI G3 Helios dual beam system. The lamellae were cut using gallium ions at 30kV and a current of 47nA for coarse and 21nA for fine milling. These lamellae were transferred from the crystal to a sapphire substrate by hand using a micromanipulator under an optical microscope. The samples were fixed using two-component araldite rapid epoxy. Low ohmic contacts were achieved by argon etching the surface at 200V for 5min before depositing 5nm of titanium and 100nm of gold. The gold film was partially removed in the FIB, exposing the region of interest of the sample, to avoid parasitic resistances. In the final step, the sample is patterned to ensure a homogenous current flow along the c-axis (figure 1A). Three different samples were prepared (two from $PtCoO_2$ and one from $PdCoO_2$). After measuring the temperature, field and in-plane angle dependence of all samples, the width of one $PtCoO_2$ structure was reduced to 2µm using a Zeiss CrossBeam FIB. All sample dimensions were measured using a scanning electron microscope and are listed in supplementary table 1.

**Resistivity Measurements.** Resistivity measurements were carried out in a 14T Quantum Design PPMS using a single axis rotator. A SynkTek multichannel lockin-amplifier was used to perform four-point resistance measurements. An alternating current of 100µA at a frequency of 177Hz was used for the measurements. This was done by sourcing a voltage to the sample in series with a current limiting resistor (1kΩ).

**Irradiation.** The electron irradiation is performed at the SIRIUS Pelletron linear accelerator operated by the Laboratoire des Solides Irradiés (LSI) at the Ecole Polytechnique in Palaiseau, France. The kinetic energy of electrons was set to 2.5 MeV, and the total electron beam current density was kept below 12µA/cm², passing through a 5 mm diameter diaphragm. During the irradiation the samples were immersed in a flow of liquid hydrogen at $T \approx 22$ K. The samples were irradiated over the course of 21 hours, to a maximum dose of 0.84 C/cm². During the irradiation the beam is swept horizontally and vertically at two incommensurate frequencies to guarantee a uniform creation of defects. All defects created by 2.5 MeV electrons hitting sample, which was cooled to 20K, are Frenkel pairs on all sublattices.

## Sample information

All sample dimensions were determined using a scanning electron microscope. The length l was determined by using the midpoint of each voltage contacts.

| Sample | w (μm) | d (μm) | l (μm) | $\rho$(2K) (μΩcm) | $\rho$(300K) (μΩcm) |
|---|---|---|---|---|---|
| PtCoO2 ac1 | 4.76 | 4.5 | 14.7 | 136 | 2284 |
| PtCoO2 ac1 thinned | 2.0 | 4.5 | 14.7 | 186 | 1064 |
| PtCoO2 ac2 | 3.4 | 4.8 | 19.5 | 138 | 2784 |
| PdCoO2 ac1 original | 5.5 | 3.9 | 22.3 | 23 | 979 |
| PdCoO2 ac1 thinned 1 | 2.7 | 3.2 | 22.3 | | |
| PdCoO2 ac1 thinned 2 | 1.7 | 3.2 | 22.3 | | |
| PdCoO2 Xirr 1 thick | 8.0 | 7.6 | 9 | 57 | |
| PdCoO2 Xirr 1 thin | 1.2 | 4.4 | 9.6 | 84 | |
| PdCoO2 ac3w | 11.9 | 4.8 | 11.1 | 82.9 | |
| PdCoO2 ac4t | 6.9 | 5.0 | 10.5 | 85.4 | |

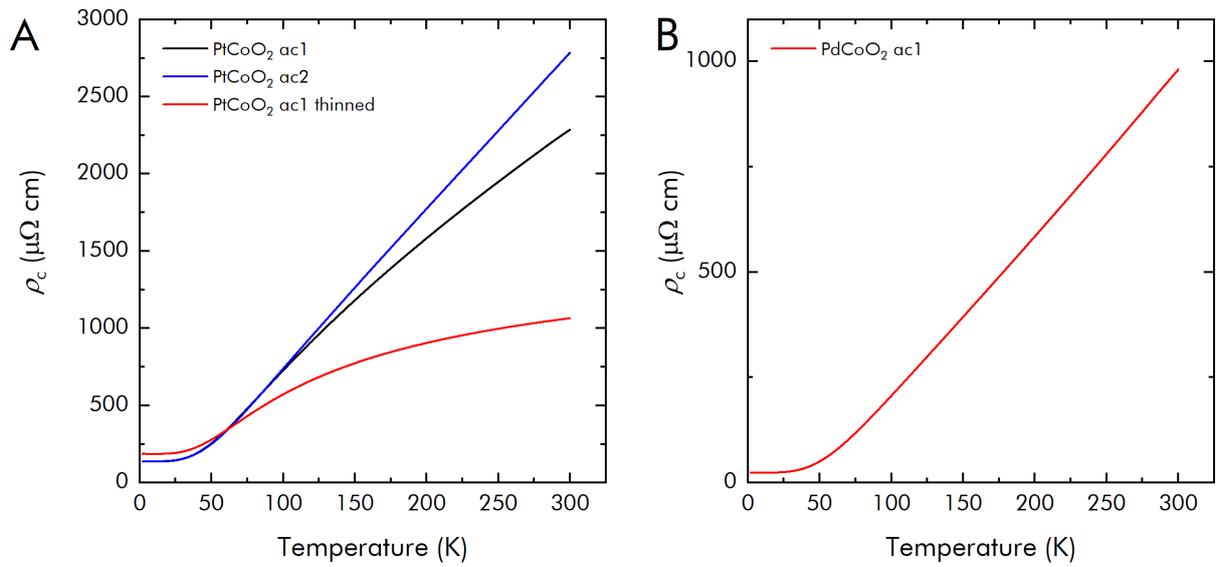

Fig. S1: Resistivity as a function of temperature for the samples used in this study. A metallic behaviour is found for the c-axis resistivity for all devices. The data on the largest devices are in good agreement with previous reports on bulk single crystals[2]. However, a strong dependence on the smallest in-plane sample dimension is observed upon thinning the samples much below the in-plane mean free path. The high temperature limit of enhanced conductivity is consistent with a parallel conduction path that does not scale with the sample cross-section but rather its circumference. A similar parallel conduction path has been reported in NbAs[3], due to the nano-meter sized FIB amorphization layer.

In the low temperature limit the smallest devices show the highest residual resistivity, opposite to the room temperature results. This reflects the enhanced boundary scattering as the sample enters the ballistic limit. Due to the high bulk conductivity the surface layer does not contribute significantly to the total conduction. The cross-over into the ballistic regime is indicated in the data by the crossing of the resistivity curves, consistent with other reports of the onset of ballistic phenomena[4]. Further evidence for this scenario is seen in the field dependence Fig. S3.

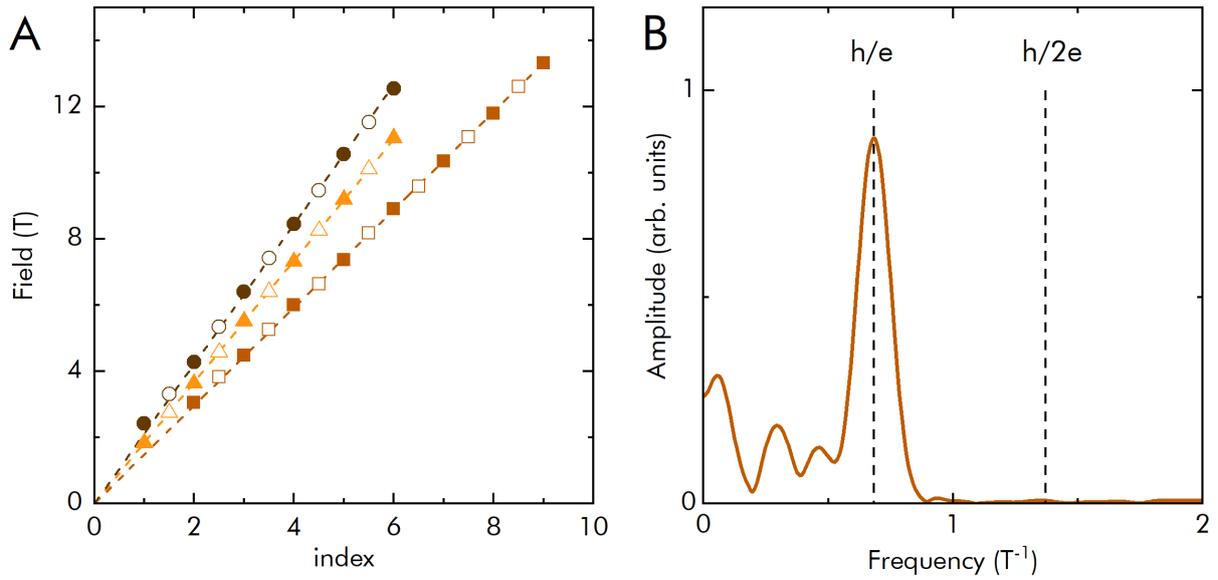

*Fig. S2: Oscillation period and phase*
A) The maxima (solid) and minima (open) positions of the phase coherent oscillations are shown versus the oscillation index for the same samples as in figure 1. (squares - PtCoO$_2$ 4.8µm; triangles – PdCoO$_2$ 3.9µm; circles – PtCoO$_2$ 3.4µm). The dashed line presents a linear in *B* dependence of the oscillation maxima and minima. The data are in good agreement with the existence of an oscillation maximum in zero magnetic field.
B) Fast Fourier Transform of the data for PtCoO$_2$ w=4.8µm (dataset shown in Fig 1C). The dashed lines indicate the frequency positions for the *h/e* and *h/2e* oscillations in a box of S=w*c/3 as illustrated in Fig 1A. If present, the amplitude of oscillations with period *h/2e* is a factor 100 times smaller than *h/e* oscillations, within the experimental noise.

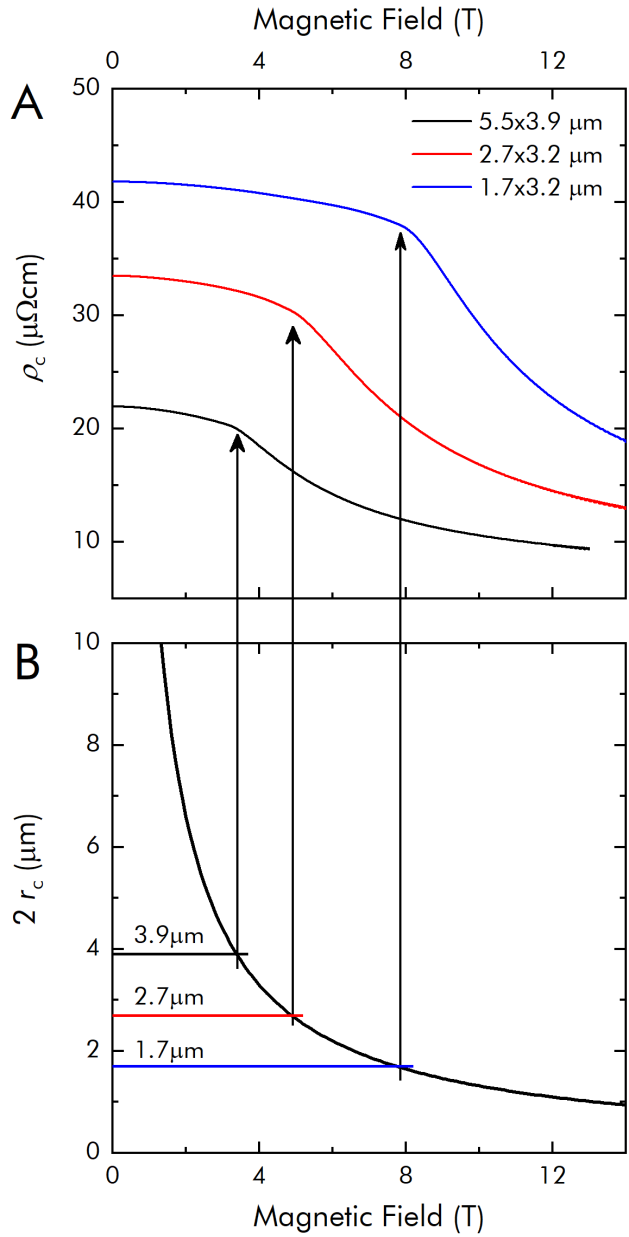

Fig. S3: Longitudinal magnetoresistance for different sample widths.
A) The magnetoresistance has been measured for a PdCoO$_2$ sample along the c-axis for different sample geometries. While the length of the bar remained unchanged, the width w and thickness d were varied as indicated in the legend. An almost field independent magnetoresistance is observed at low field which is followed by a pronounced drop in magnetoresistance at high magnetic field.

B) The cyclotron diameter $2r_c = 2\hbar k_F/eB$ with the maximum $k_F = 1\text{Å}^{-1}$ (solid black line) value of the Fermi wave vector of the hexagonal Fermi surface[5]. Lines and arrows indicate the field strength at which the cyclotron diameter becomes smaller than the lesser of the two in-plane dimensions.

This demonstrates that the increase in conductivity is due to the suppression of surface scattering as the sample transitions from the mesoscopic to the bulk limit

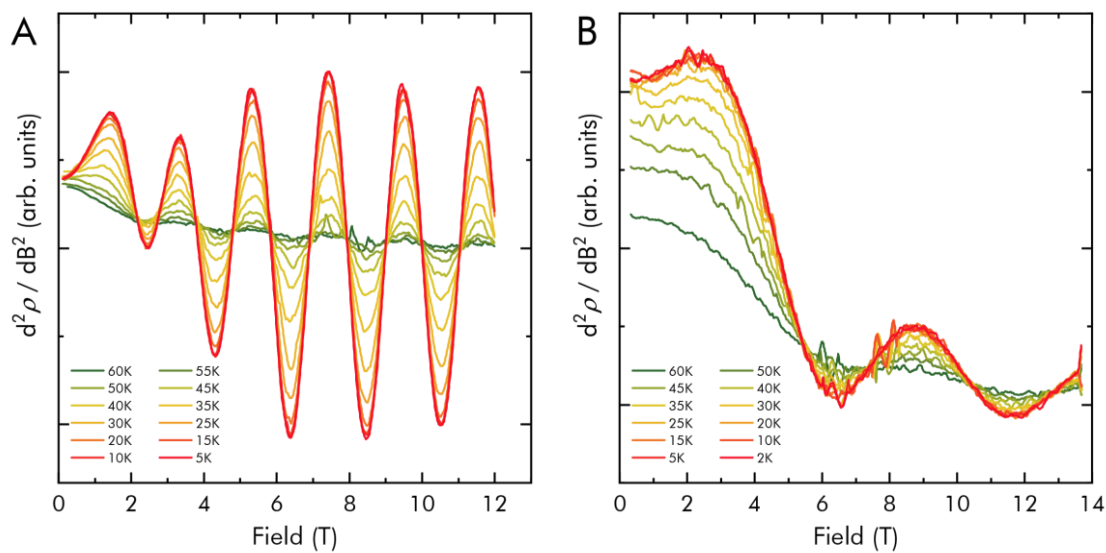

Fig. S4: Supporting data for Figure 4 of the main manuscript.
A) Second derivative of magnetoresistance with respect to magnetic field for $PtCoO_2$ $w=4.8\mu m$.
B) Second derivative of magnetoresistance with respect to magnetic field for irradiated $PdCoO_2$ $w=1.1\mu m$.

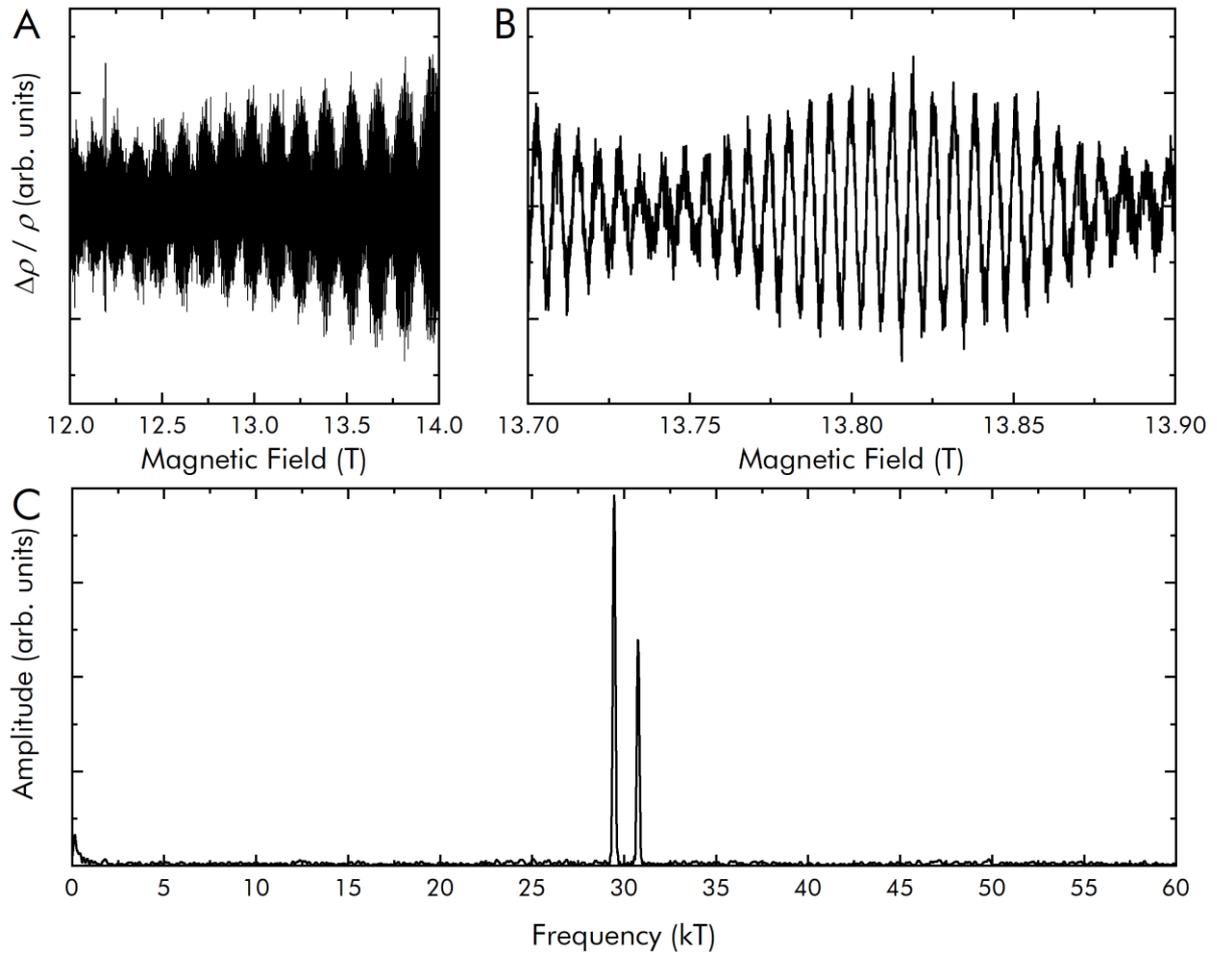

Fig. S5 Quantum oscillations periodic in 1/B for the sample PdCoO$_2$ Xirr1 thick. The measurement was performed for the longitudinal configuration with magnetic field and current applied along the c-axis of the sample. Panel A shows the signal after a polynomial background was subtracted. Panel B presents a zoom of the oscillations. The dominant frequencies of 30kT shown in the FFT in panel C are in good agreement with the Fermi surface observed in angle resolved photo emission spectroscopy[6].

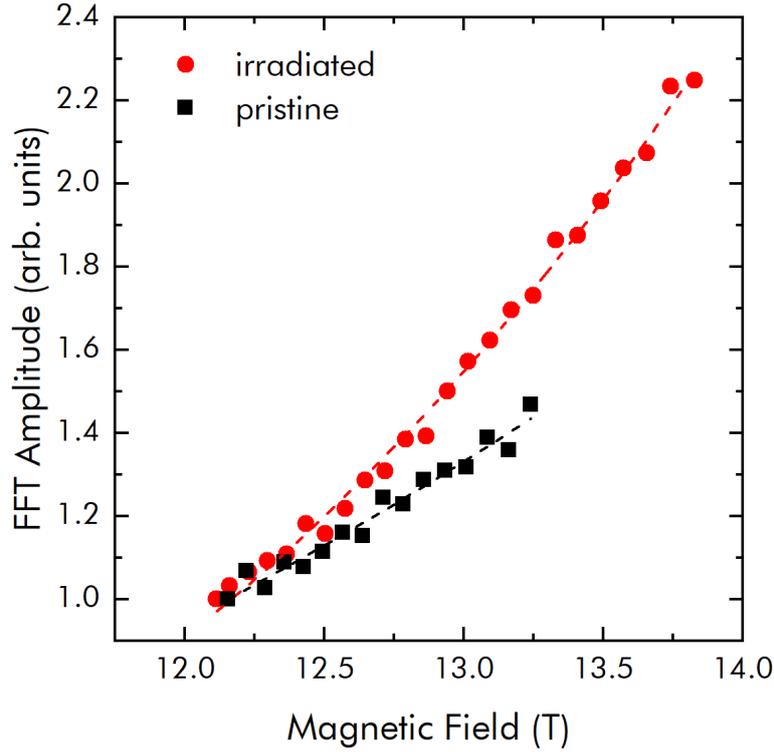

Fig S6: Quantum oscillation amplitude of different microstructures of pristine and irradiated samples of $PdCoO_2$.

To determine the quantum coherent mean free path from SdH oscillations we have performed FFT in field windows of constant width in $1/B$ space. The sliding window was chosen as $\Delta 1/B = 1.8\text{e-}3$ $T^{-1}$ such that the two frequencies at 30kT and 31kT were resolved in the FFT. The extracted amplitudes are normalized at the lowest field value and fitted by the Dingle term of the Lifshitz-Kosevich formula, $A(B) = c * \exp(-\alpha/B)$. From this a $k$-independent quantum coherent mean free path can be extracted by assuming a cylindrical Fermi surface by $l_{qc} = 1140 * \sqrt{F}/\alpha$, with the oscillation Frequency $F$. We find $l_{qc} = 373nm$ and $l_{qc} = 240nm$ respectively for the pristine and irradiated sample, which is a small change compared to the order of magnitude change found in the transport mean free path.

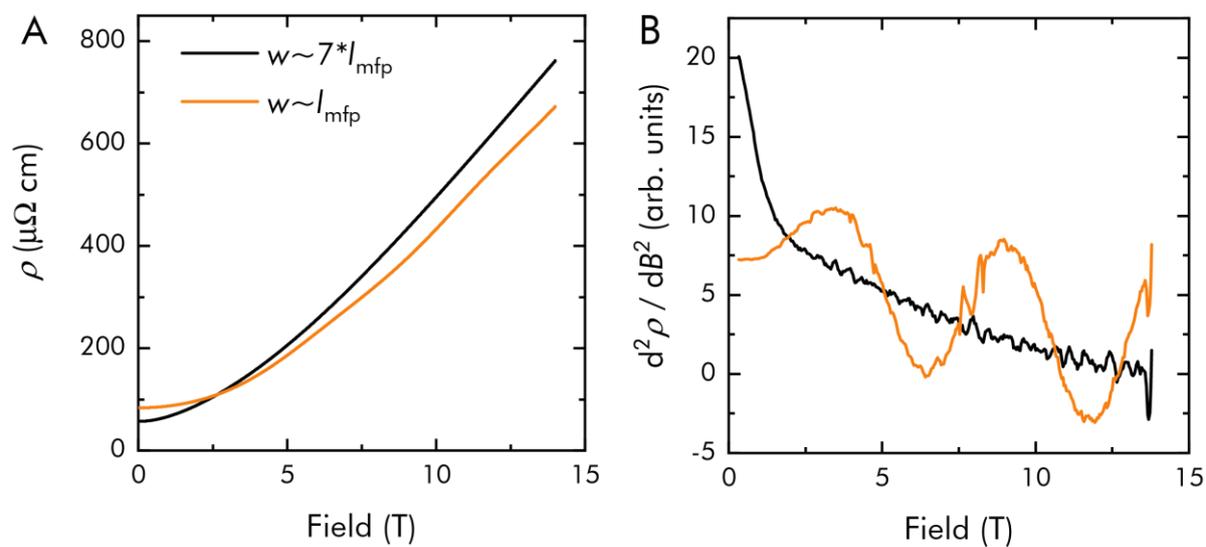

Fig. S7: A) Magnetoresistance of c-axis of irradiated $PdCoO_2$. The mfp was determined from in-plane resistivity measurements (see suppl. figure 1) to be 1μm. The sample width was varied from 8μm (black) to 1μm (orange).
B) Second derivative of magnetoresistance with respect to magnetic field for irradiated $PdCoO_2$. The *h/e*-oscillations are observed for sample that whose width is of the order of the mfp in the Pd-layer.

## Model calculation of interlayer conductance

We perform a microscopic calculation of the interlayer conductance using the finite system Kubo formula. The model we employ is the anisotropic Hofstadter model (figure 5A), a tight binding model of electrons on a two dimensional square lattice with hopping parameters $\tau_1$ for in-plane and $\tau_\perp$ for interlayer tunnelling set to $\tau_1 = 1$, $\tau_\perp = 0.01\tau_1$. We impose open boundary conditions in the in-plane direction and label the sites as $j=1...L$, while periodic boundary condition is imposed on the interlayer direction. The magnetic field is introduced via a gauge field $A_\perp = 2\pi\frac{\varphi}{L} j$ making the interlayer tunnelling acquire a site dependent phase modulation $\tau_\perp e^{i 2\pi\frac{\varphi}{L} j}$ as outlined in the main text. The field strength is written such that there are $\varphi = |\vec{B}\cdot\vec{S}_i|/\Phi_0$ flux quanta per layer across the system of width $L$. Lattice translational invariance in the interlayer direction is present and the corresponding plane wave solution is characterized by a continuous momentum $k_z$. The Hamiltonian can be represented by a matrix with elements $H_{j+1,j} = H_{j,j+1} = -\tau_1, H_{j,j} = -2\tau_\perp \cos(k_z + 2\pi\frac{\varphi}{L} j)$ and zero otherwise. By diagonalizing this matrix, we obtain eigenstates $|n\rangle$ $n=1...L$ with energy $\mathcal{E}_n$. We place the Fermi energy at $E_F = 0$.

Since the model is finite in the in-plane direction, we can consider it as a one-dimensional $L$-band system extended in the interlayer direction. The energy dispersion of bands near the Fermi energy is plotted in figure 5 B of the main text. While they are dispersive at zero field, the band width narrows and becomes zero when the flux approaches positive integers ($\varphi \sim 1,2,...$). When this happens, interlayer transport is suppressed. To examine this, we evaluate the interlayer conductivity using the Kubo formula

$$\sigma_{\text{inter}} = \frac{ie^2}{L} \sum_{n,m=1\,(n\neq m)}^{L} \int_{-\pi}^{\pi} \frac{dk_z}{2\pi} \left(\frac{f(\mathcal{E}_n) - f(\mathcal{E}_m)}{\mathcal{E}_m - \mathcal{E}_n}\right) \frac{|\langle n|j_z|m\rangle|^2}{\mathcal{E}_n - \mathcal{E}_m + i\eta_1}$$

$$\sigma_{\text{intra}} = \frac{e^2}{\eta_2} \frac{1}{L} \sum_{m=1}^{L} \int_{-\pi}^{\pi} \frac{dk_z}{2\pi} \left(-\frac{\partial f(\mathcal{E}_m)}{\partial \mathcal{E}}\right) |v_n^z|^2.$$

We have expressed the inter-band ($n \neq m$) and intra-band ($n = m$) contributions separately. The interlayer current operator is represented by a diagonal matrix $(j_z)_{j,j} = 2\tau_\perp \sin(k_z + 2\pi\frac{\varphi}{L} j)$, and the Fermi velocity is denoted as $v_n^z = \langle n|j_z|n\rangle$. We have introduced two phenomenological scattering parameters $\eta_1$ [28,29] and $\eta_2$. Choosing $\eta_1 \sim 1/L^2$ yields an interband contribution to the conductivity that is finite. The resulting conductivity is plotted in the inset of figure 5 C of the main text, where we have used $\eta_1 = 1/L^2$, $\eta_2 = 0.1$. The interlayer resistivity is obtained by inverting the conductivity $\rho = 1/(\sigma_{\text{inter}} + r\,\sigma_{\text{intra}})$, where we introduced a parameter $r$ which changes the ratio of the two contributions. In fact, this parameter is redundant and we can consider $\eta_2/r$ as the overall scattering parameter of the intra-band contribution. As we increase the flux from zero, the intra-band contribution $\sigma_{\text{intra}}$ becomes strongly suppressed and approaches zero around the positive integers $\varphi \sim 1,2,...$. The inter-band contribution $\sigma_{\text{inter}}$ also shows an oscillation and is weakly peaked around the positive integers, rather than showing dips as in $\sigma_{\text{intra}}$. The resistivity plotted in figure 5 C inherits this oscillation ($r = 0.0075$ is used).

How can we understand the oscillation? One possible interpretation is the effect of multi-path interference. When the ratio $\tau_\perp/\tau_1$ is small, a state $|n\rangle$ near the Fermi energy is a standing wave extended over the layer, i.e. $|\psi^n(j)|^2 \sim 1/L$, where $\psi^n(j)$ is the real space wave function. Interlayer transport takes place when an electron in this extended state tunnels to an adjacent layer. This happens through all $L$ possible paths, each having a phase modulation of $e^{i\,2\pi\frac{\varphi}{L}j}$. Using the tunnelling amplitude $A(\varphi) = \sum_{j=1}^{L} e^{i\,2\pi\frac{\varphi}{L}j}$ introduced in the main text, the group velocity behaves as $v_n^z \sim \frac{\tau_\perp}{L}\,Im\,A(\varphi) \sim \tau_\perp \frac{1-\cos(2\pi\varphi)}{2\pi\varphi}$ and shows an $\varphi \sim \varphi + 1$ oscillation with an overall decay. This explains the decrease and oscillation of the interlayer conductivity since $\sigma_{\text{intra}}$ is roughly proportional to $(v_n^z)^2$ in this limit.

## Effect of enhanced edge hopping

As mentioned in the main text, the most naïve picture of a simple Aharonov-Bohm loop would be one in which hopping at the edges of the sample dominated that in the bulk. In order to examine this possibility, we consider a model with an enhanced interlayer hopping $\tau'_\perp$ at the two edges ($\tau'_\perp = 0.05\tau_1$, $\tau_\perp = 0.01\tau_1$,), as sketched in Panel A of Fig. S1. Although the calculation gives oscillations of the correct period, they die off as $1/L$ and would therefore be completely invisible in samples such as ours for which $L \sim 10^4$.

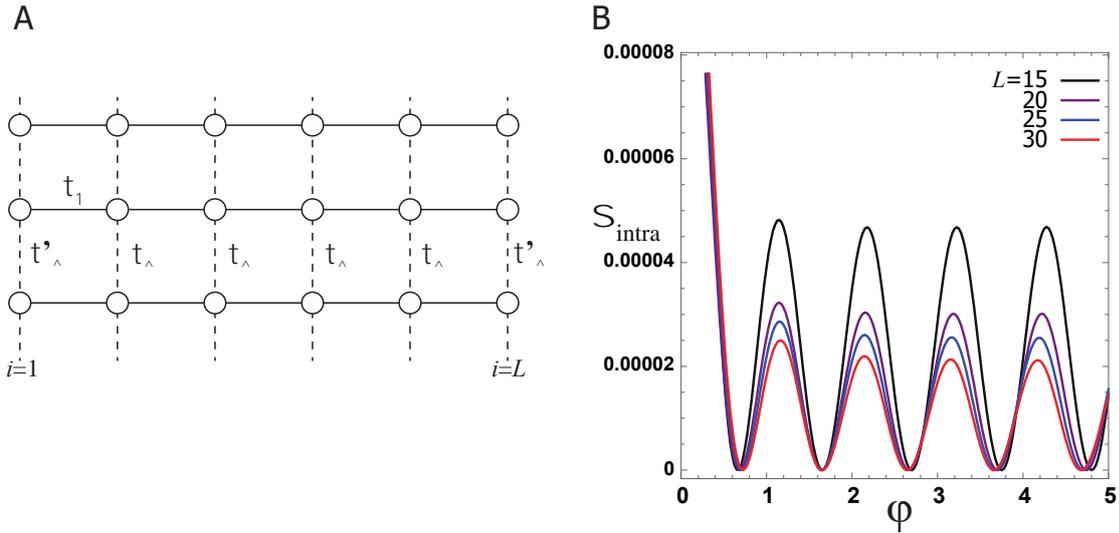

Fig. S8. A Schematic illustrating a situation in which edge hopping $\tau'_\perp$ is larger than that in the bulk. B Results for the intra-layer conductivity, in which the resultant oscillations die off as $1/L$.

Fig. S9 Analysis of Data for in-plane rotation of PdCoO2 ac1

In the following pages the analysis for the B-linear oscillation is demonstrated based on the in-plane angle dependence of *PdCoO2 ac1 original*. The results of this analysis are shown in Fig.2B.

Left panels show the raw magnetoresistance. From this the second derivative with respect to magnetic field is shown in the middle panels. The data was normalized to the magnetoresistance background. The data was further normalized to its maximum value in the field range of 2 to 14T for clarity. To extract the oscillation frequency a FFT is performed of the data in the range of 2 to 14T. The low field range was not included in the analysis as the initial sharp increase in magnetoresistance causes a stark peak at zero frequency hindering the analysis of low frequency oscillation. The resulting frequency spectrum is shown in the right panel. The identified peaks in the spectrum are labelled. For each line the magnetic field angle as shown in Fig.2 is indicated.

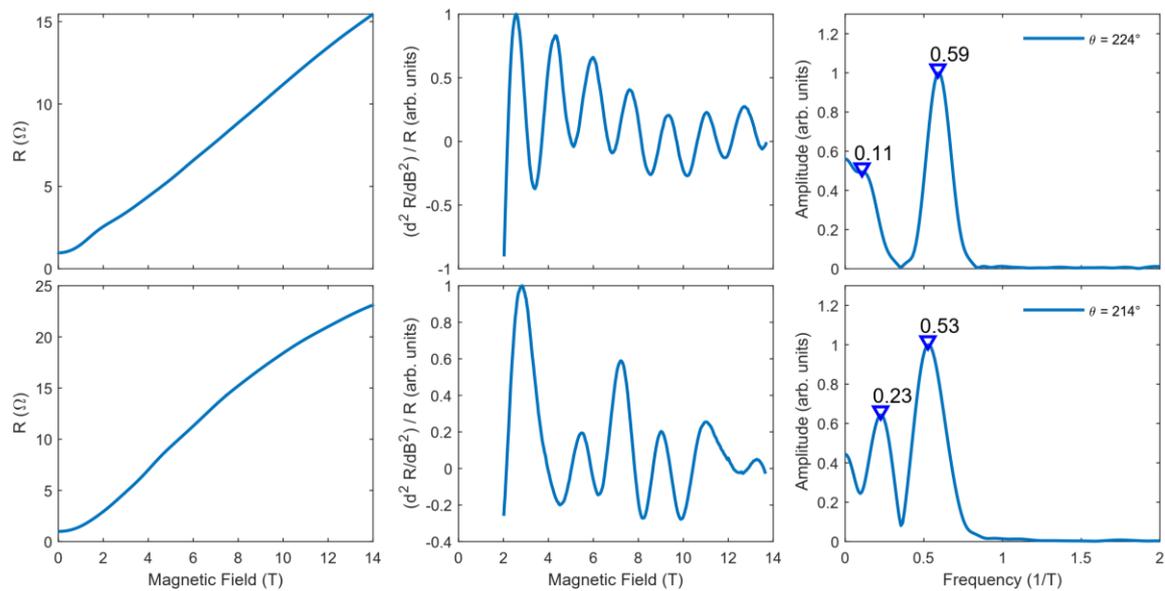

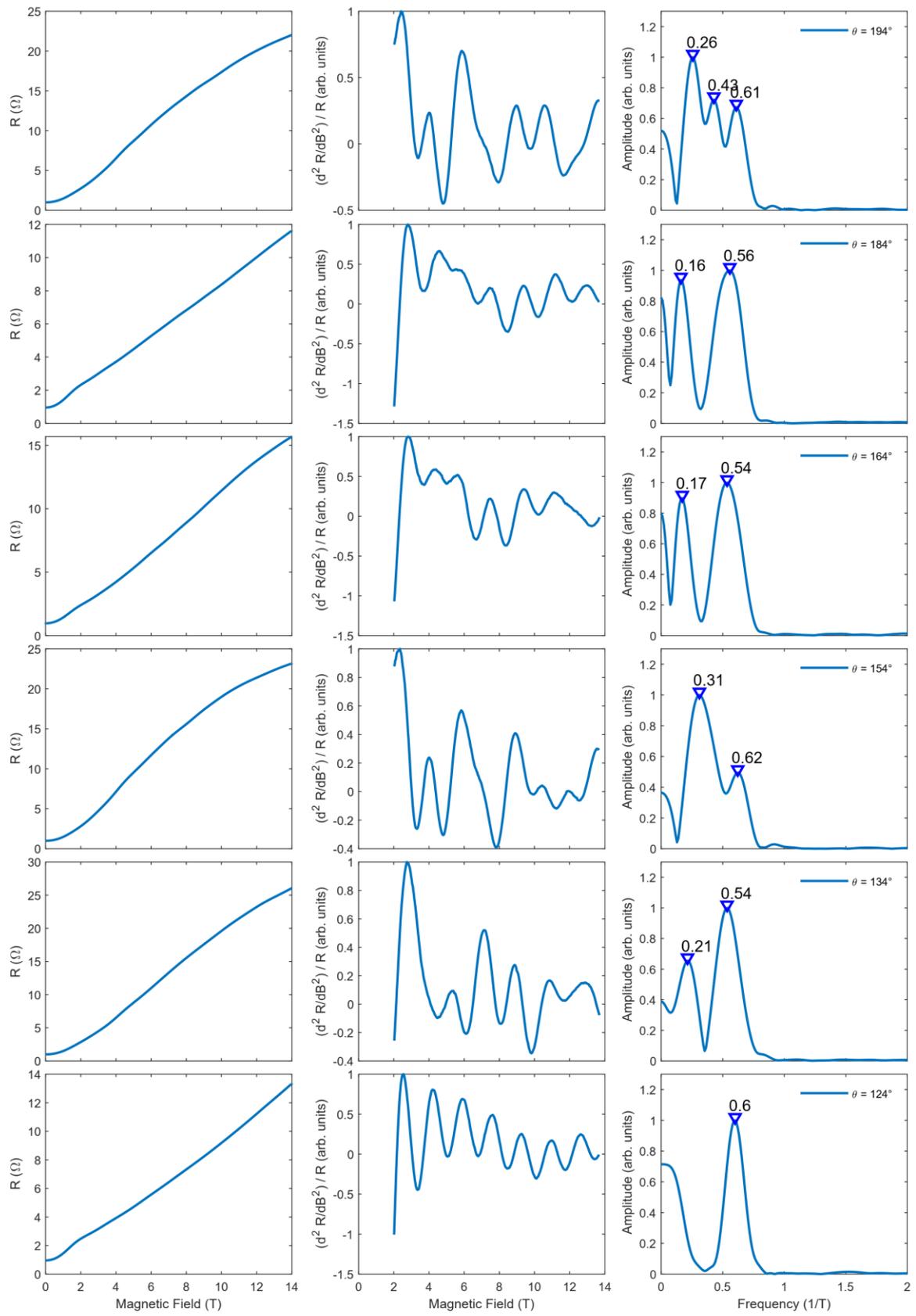

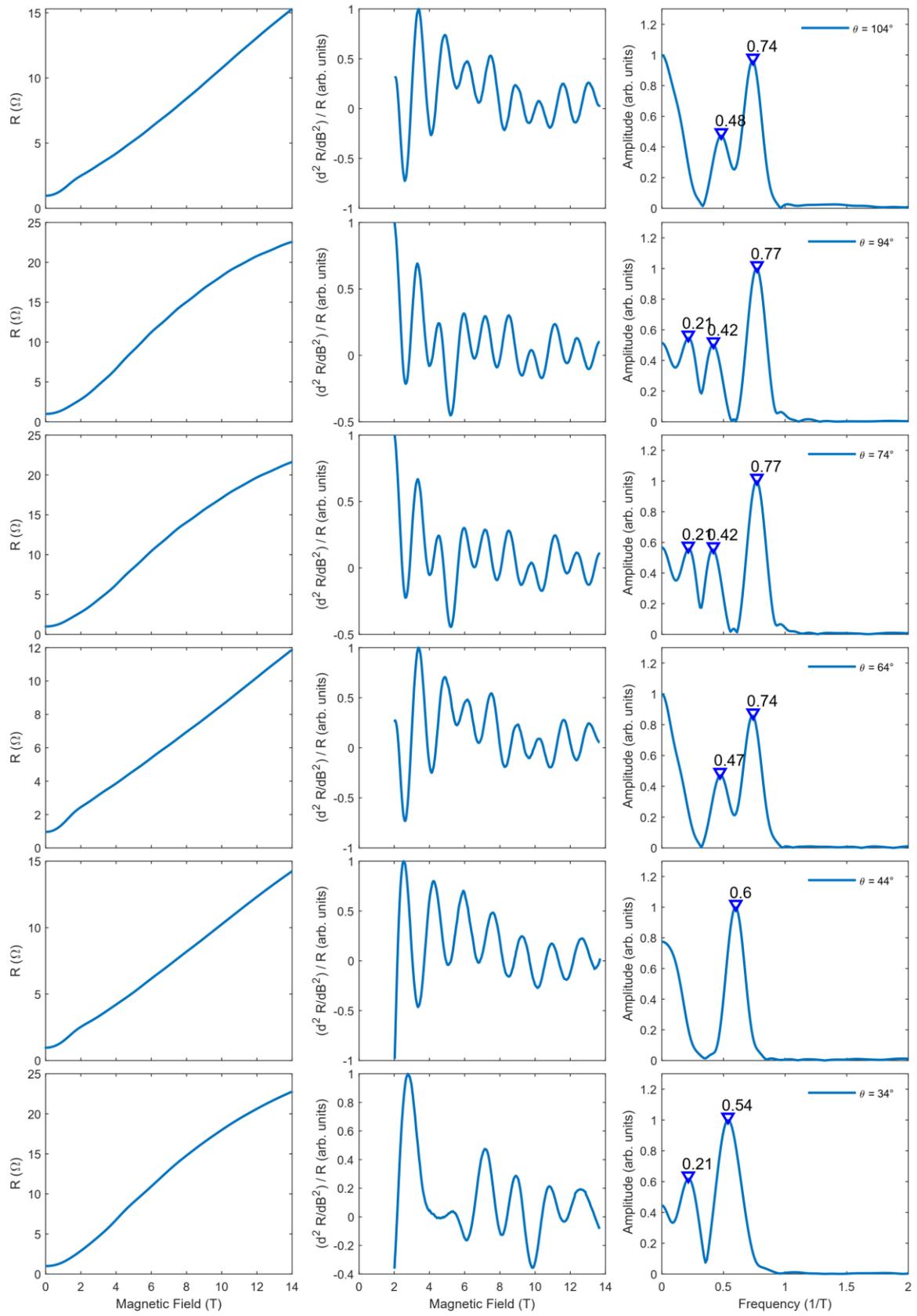

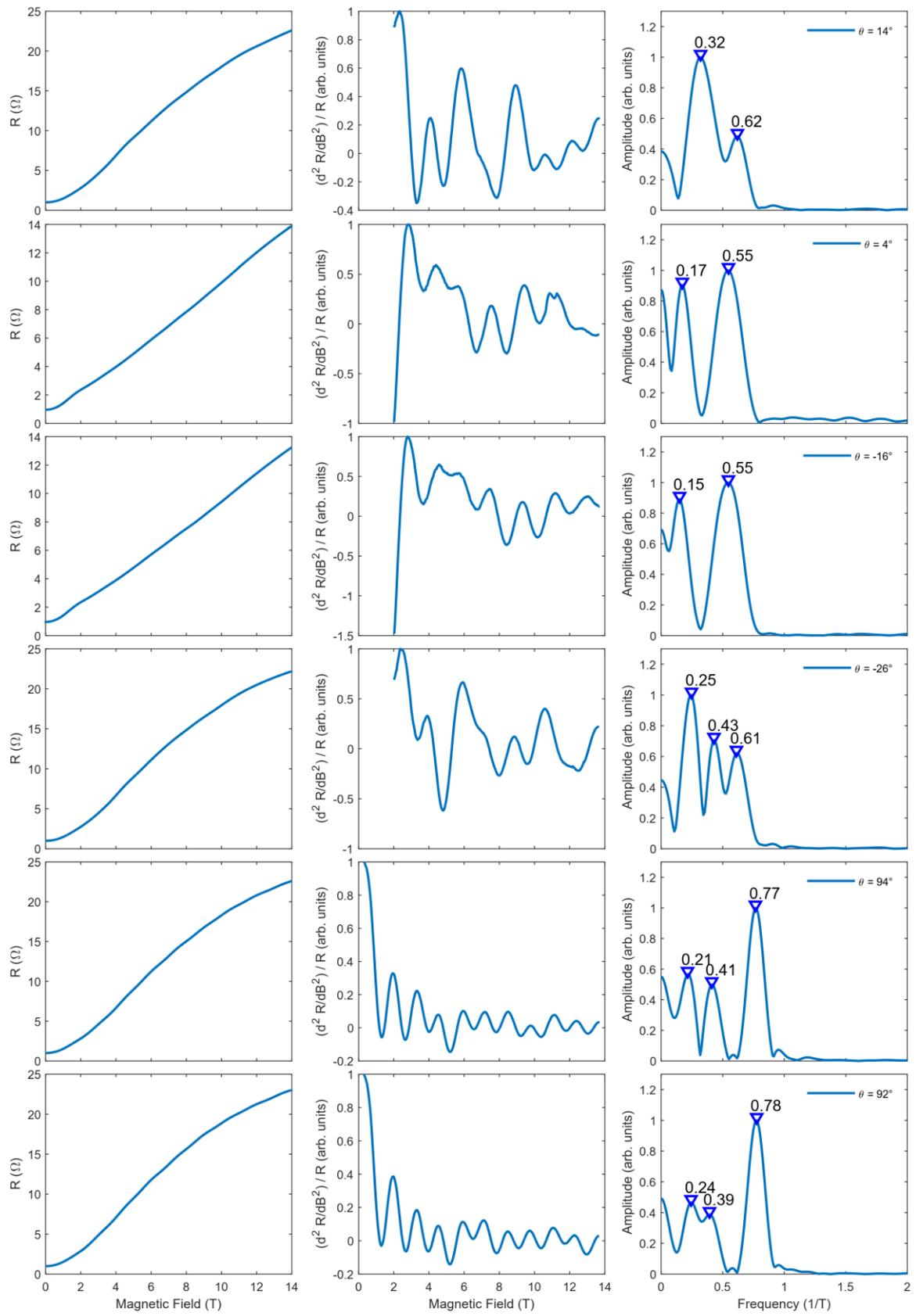

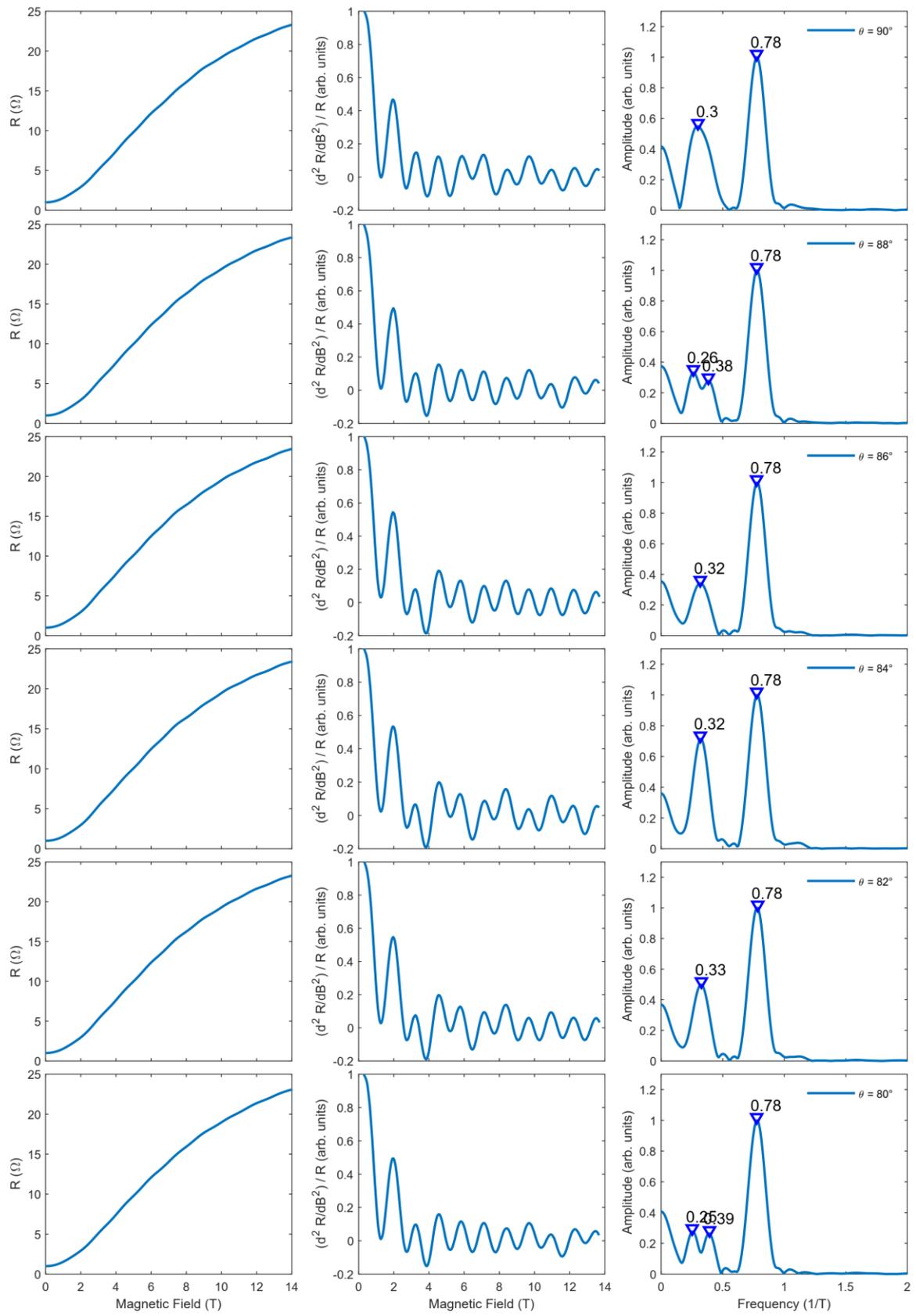

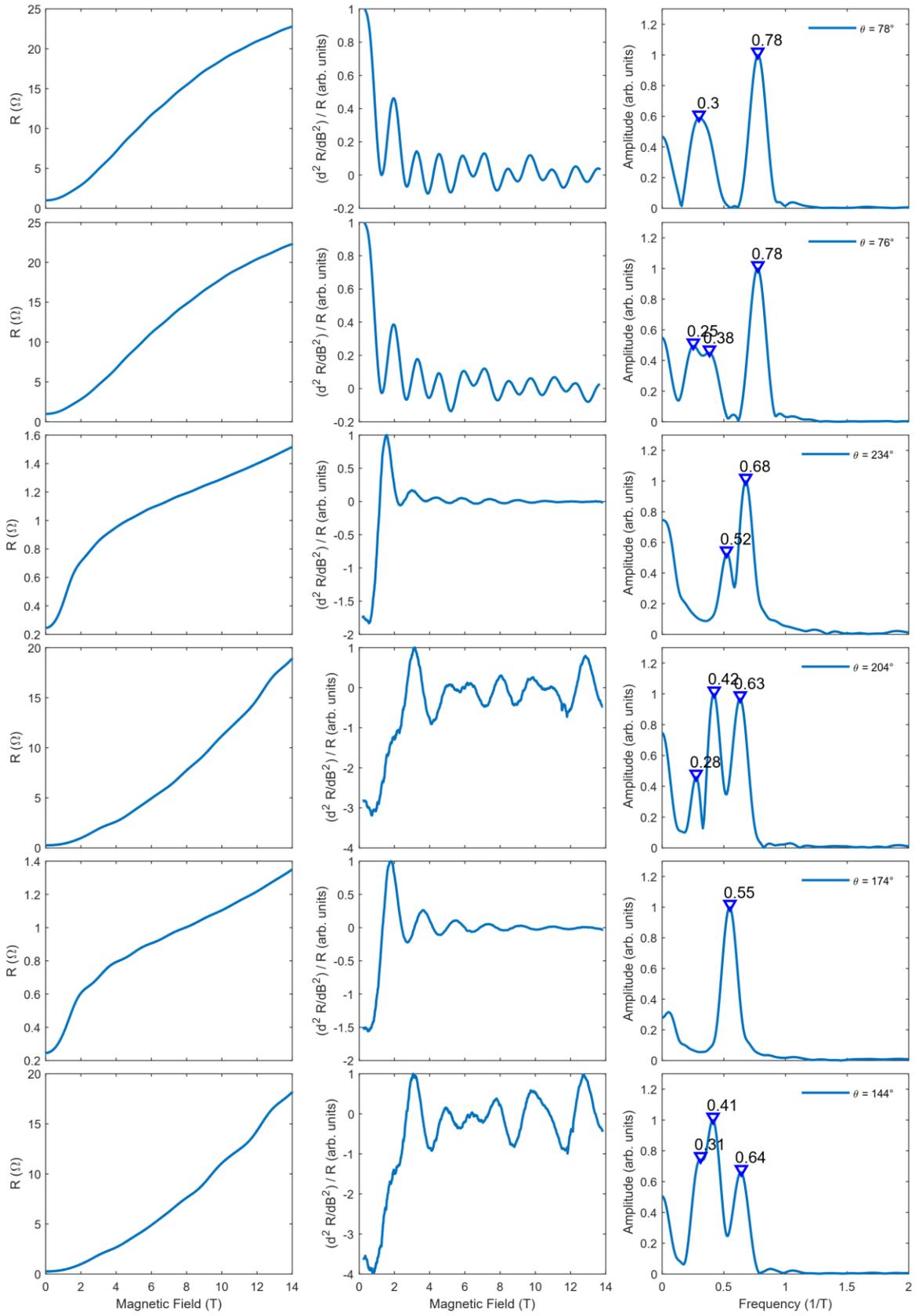

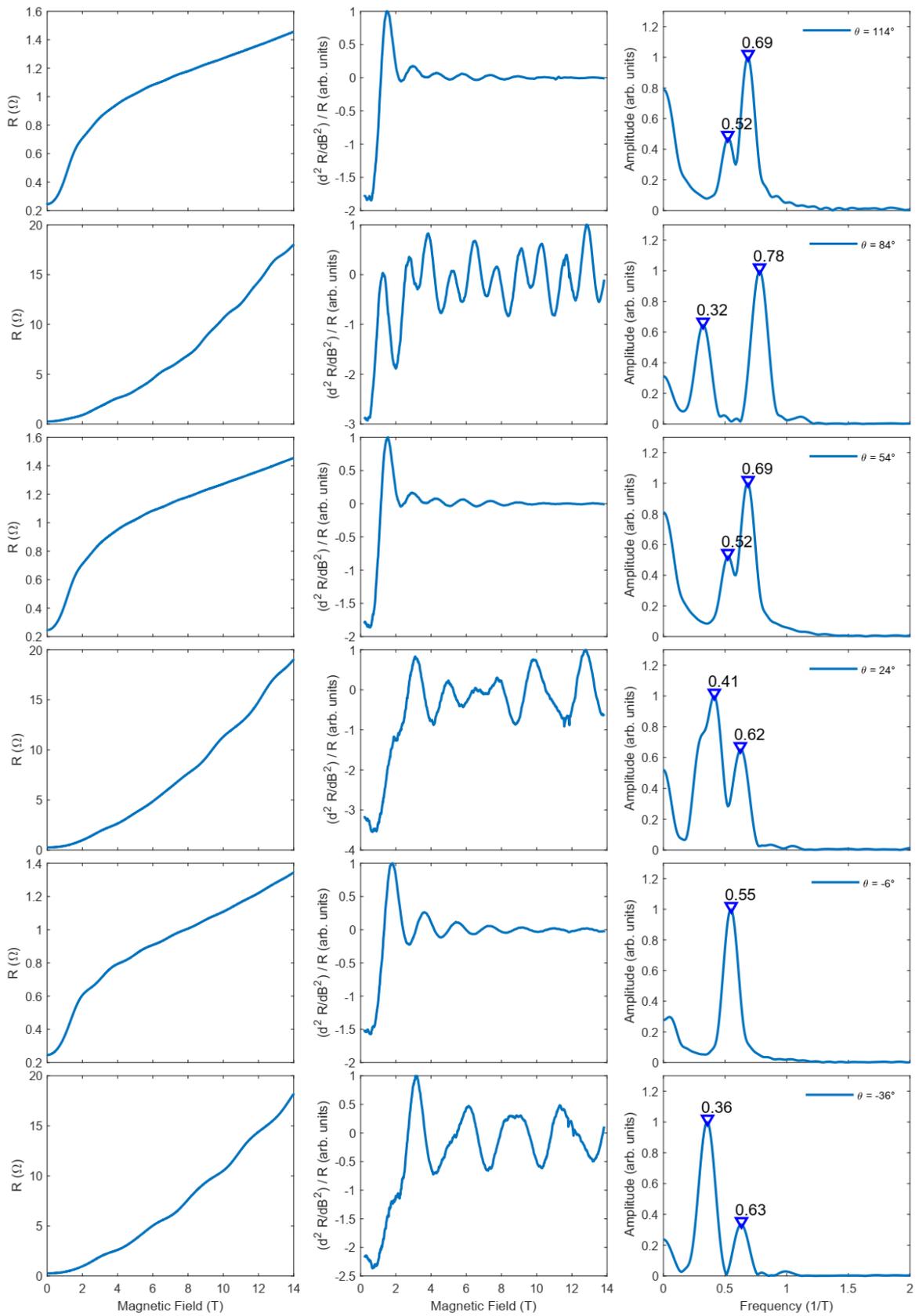